\documentclass[manuscript]{aastex}

\shorttitle{N K-shell photoabsorption}
\shortauthors{Garc\'{\i}a et al.}

\begin{document}


\title{Nitrogen K-shell photoabsorption}


\author{J. Garc\'ia,}
\affil{The Catholic University of America, IACS, Physics Department, Washington DC 20064; and
       NASA Goddard Space Flight Center, Greenbelt, MD 20771}
\email{javier@milkyway.gsfc.nasa.gov}

\author{T.R. Kallman, M. Witthoeft, E. Behar\footnotemark}
\affil{NASA Goddard Space Flight Center, Greenbelt, MD 20771}
\email{michael.c.witthoeft@nasa.gov;
     timothy.r.kallman@nasa.gov; behar@milkyway.gsfc.nasa.gov}

\author{C. Mendoza,}
\affil{Centro de F\'isica, IVIC, Caracas 1020A, Venezuela}
\email{claudio@ivic.ve}

\author{P. Palmeri,}
\affil{Astrophysique et Spectroscopie, Universit\'e de Mons,
      B-7000 Mons, Belgium}
\email{palmeri@umons.ac.be}

\author{P. Quinet,}
\affil{Astrophysique et Spectroscopie, Universit\'e de Mons, B-7000 Mons,
  Belgium, and
\\ IPNAS, Sart Tilman B15, Universit\'e de Li\`ege, B-4000 Li\`ege, Belgium}
\email{quinet@umons.ac.be}

\author{M.A. Bautista\footnotemark}
\affil{Virginia Polytechnic Institute and State University, Blacksburg, VA 24061}
\email{bautista@vt.edu}

\author{and M. Klapisch}
\affil{ARTEP, Inc., Ellicott City, MD 21042, USA}
\email{marcel.klapisch.ctr@nrl.navy.mil}


\altaffiltext{1}{Visiting senior NPP fellow. Permanent address: Department of Physics, Technion, Haifa 32000, Israel}
\altaffiltext{2}{Present address: Department of Physics, Western Michigan University, Kalamazoo, MI 49008}



\begin{abstract}
Reliable atomic data have been computed for the spectral modeling of the nitrogen
K lines, which may lead to useful astrophysical diagnostics. Data sets comprise
valence and K-vacancy level energies, wavelengths, Einstein $A$-coefficients,
radiative and Auger widths and K-edge photoionization cross sections. An important
issue is the lack of measurements which are usually employed to fine-tune calculations
so as to attain spectroscopic accuracy. In order to estimate data quality, several
atomic structure codes are used and extensive comparisons with previous theoretical data
have been carried out. In the calculation of K photoabsorption with the
Breit--Pauli $R$-matrix method, both radiation and Auger damping, which cause the
smearing of the K edge, are taken into account. This work is part of a wider project
to compute atomic data in the X-ray regime to be included in the database of the
popular {\sc xstar} modeling code.

\end{abstract}






\section{Introduction}

The improved resolution and sensitivity of current satellite-borne
X-ray observatories ({\em Chandra} and {\em XMM Newton}) are allowing the study of
previously inaccessible weak spectral features of astrophysical interest.
In the early stages of these missions, it was realized that absorption by
near-neutral species was common, and the fact that all charge states (except
the fully ionized) left identifiable imprints in the X-ray spectrum has proven to
be a powerful diagnostic. Inner-shell absorption is important in the outflows of
Seyfert galaxies in terms of both Fe L$\alpha$ \citep{sako01, behar01} and K$\alpha$ lines
of high-Z elements \citep{behar02}, and also of elements in the first row of the
periodic table such as oxygen \citep{pradhan00, behar03, gar05}. 
Furthermore, inner-shell absorption of continuum X-rays from bright
galactic sources is a useful diagnostic of the interstellar medium
\citep{yao09, kaastra09}.

Nitrogen K-shell absorption and emission are detected in X-ray spectra,
mostly due to the H- and He-like charge states. For instance, observations of
the emission lines of \ion{N}{6} and \ion{N}{7} in the ejecta of $\eta$~Carinae by
\citet{leu03} have resulted in the lower bound ${\rm N/O}>9$ for its
nitrogen abundance. This result puts a constraint in the evolution of $\eta$~Car
and is a signature of CNO-cycle processing. The \ion{N}{6} triplet is
observed in the spectrum of the magnetic B star $\beta$~Cep where it has been used
to test magnetically confined wind shock models \citep{fav09}. It is found that
the plasma is not heated by magnetic reconnection and there is no evidence for an
optically thick disk at the magnetic equator. Highly ionized emission lines of
nitrogen have been observed by \citet{miy08} on the north-eastern rim of the Cygnus
Loop supernova remnant which can be used to determine nitrogen 
abundances, which turn out to be 23\% solar.
Seyfert galaxy outflows also can have super-solar N abundances \citep{brinkman02, arav07}.

Narrow absorption K$\alpha$ and K$\beta$ lines of \ion{N}{6} have been
identified in the warm absorber of the MR 2251-178 quasar, which point to a complex
velocity field with an outflow of ionized material \citep{ram08}. K-shell absorption
by Li-like \ion{N}{5}, which has prominent UV lines, or by lower charge states at
wavelengths $\lambda > 29$~\AA\, is difficult to detect due to the reduced sensitivity
of current X-ray instruments towards these longer wavelengths.
Notable exceptions include : K$\alpha$
absorption by \ion{N}{5} at 29.42~\AA\ has been reported by \citet{steenbrugge05} in
the outflow of NGC~5548; and N absorption by a white dwarf outflow
has been observed following the outburst of nova V4743 Sagittarii \citep{ness03}. In the
latter case, only the H- and He-lines are discussed, but lower charge states of N are
clearly seen in the spectrum longward of 29~\AA\ \citep[see Fig.~3b in][]{ness03}.

The current proliferation of X-ray spectra with high signal-to-noise ratio in astronomical
archives makes the computation of nitrogen K-shell photoabsorption particularly timely. This
is also an additional and important step in the ongoing effort to compute reliable atomic data
by \citet{gar05, pal08a, pal08b, wit09} for K-line analysis within the context of
the {\sc xstar} spectral modeling code \citep{kal01}. Available spectroscopic data for K-vacancy
levels of the N isonuclear sequence are notably scarce, mainly limited to \ion{N}{5} and \ion{N}{6} for
which a few levels are listed in the NIST database \citep{ral08} and four measured wavelengths
have been reported by \citet{bei99}. This shortage of reliable measurements precludes
empirical corrections to calculated level energies. On the other hand, several relativistic
methods have been previously used to generate atomic data for nitrogen K-vacancy states:
the saddle-point complex-rotation method \citep{dav89,chu90,shi01,lin01,lin02,zha05}; the Raleigh--Ritz
variational method \citep{hsu91, yan95,wan06}; and multiconfiguration Dirac--Fock (MCDF) approaches
such as those by \citet{hat83,har84,che86,che87,che88,che97}. Photoabsorption cross sections in the
near K-edge region of N ions have been obtained by Hartree--Slater central-field computations
\citep{rei79}, where the resonance structure due to quasibound states as well as
configuration correlations are neglected. The net effect of the resonance structure is to
fill in the photoionization cross section below the inner-shell threshold altering the shape
of the K edge. 

In this paper we report on calculations of energy level structure and bound-bound and bound-free
transition probabilities for the K-shell of nitrogen. 
The outline of the present report is as follows. The numerical
methods are briefly described in \S~2 while an analysis of the
results based on comparisons with previous experimental and
theoretical values is carried out in \S~3. The two
supplementary electronic tables are explained in \S~4 while
some conclusions are finally discussed in \S~5.

\section{\label{numerical_methods} Numerical methods}

The numerical approach used here has been fully described in \citet{bau03}.
Level energies, wavelengths, $A$-coefficients, and radiative and Auger
rates are computed with the codes {\sc autostructure} \citep{eis74, bad86, bad97}
and {\sc hfr} \citep{cow81}. For consistency, configuration-interaction (CI)
wave functions of the type
\begin{equation}
  \Psi = \sum c_i\phi_i
\end{equation}
are calculated with the relativistic Breit--Pauli Hamiltonian
\begin{equation}
  \label{hbp}
  H_{\rm bp} = H_{\rm nr} + H_{\rm 1b} + H_{\rm 2b}
\end{equation}
where $H_{\rm nr}$ is the usual non-relativistic Hamiltonian. The
one-body relativistic operators
\begin{equation}
   \label{h1b}
   H_{\rm 1b} = \sum_{n=1}^{N} {f_n({\rm mass}) + f_n({\rm d}) + f_n({\rm so})}
\end{equation}
represent the spin--orbit interaction, $f_n({\rm so})$, the
non-fine-structure mass variation, $f_n({\rm mass})$, and the
one-body Darwin correction, $f_n({\rm d})$. The two-body Breit
operators are given by
\begin{equation}
   \label{h2b}
   H_{\rm 2b} = \sum_{n<m} g_{nm}({\rm so}) + g_{nm}({\rm ss}) + g_{nm}({\rm css})
   + g_{nm}({\rm d}) + g_{nm}({\rm oo})
\end{equation}
where the fine-structure terms are $g_{nm}(so)$ (spin--other-orbit
and mutual spin-orbit), $g_{nm}(ss)$ (spin--spin), and the
non-fine-structure counterparts $g_{nm}(css)$ (spin--spin
contact), $g_{nm}(d)$ (two-body Darwin), and $g_{nm}(oo)$
(orbit--orbit). It must be pointed out that {\sc hfr} neglects contributions from the
two-body term $H_{\rm 2b}$ of Equation~(\ref{h2b}).

In {\sc hfr}, core-relaxation effects (CRE) are always taken into account since each electron
configuration is represented with its own set of non-orthogonal orbitals optimized by
minimizing the average configuration energy. In {\sc autostructure}, on the other hand,
configurations may be represented with either orthogonal or non-orthogonal orbitals which
then enables estimates of the importance these effects. In the present calculation five approximations are
considered in order to study the effects of electron correlation, i.e. configuration interaction (CI)
and CRE, and thus to estimate data accuracy.
\begin{description}
\item[Approximation AS1] Atomic data are computed with {\sc autostructure} including CI
from only the $n=2$ complex. CRE are neglected.
\item[Approximation AS2] Atomic data are computed with {\sc autostructure} including both $n=2$ CI
and CRE.
\item[Approximation AS3] Atomic data are computed with {\sc autostructure} including CI
from both the $n=2$ and $n=3$ complexes. CRE are neglected.
\item[Approximation HF1] Atomic data are computed with {\sc hfr} including CI
only from the $n=2$ complex.
 \item[Approximation HF2] Atomic data are computed with {\sc hfr} including CI
from both the $n=2$ and $n=3$ complexes.
\end{description}

Photoabsorption cross sections are obtained with the codes Breit-Pauli R-Matrix {\sc bprm} \citep{ber87,sea87} and
{\sc hullac} \citep{hullac01}.
In {\sc bprm}, wave functions for states of an $N$-electron target
and a colliding electron with total angular momentum and parity
$J\pi$ are expanded in terms of the target eigenfunctions
\begin{equation}\label{cc}
  \Psi^{J\pi}={\cal A}\sum_i \chi_i{F_i(r)
  \over r}+\sum_jc_j\Phi_j\ .
\end{equation}
The $\chi_i$ functions are vector coupled products of the target
eigenfunctions and the angular components of the incident-electron
functions; $F_i(r)$ are the radial part of the continuum
wave functions that describe the motion of the scattered electron;
and $\mathcal{A}$ is an antisymmetrization operator. The functions
$\Phi_j$ are bound-type functions of the total system constructed
with target orbitals. The Breit--Pauli relativistic version has
been developed by \citet{sco80} and \citet{sco82}, but the
inclusion of the two-body terms (see Equation~\ref{h2b}) is
currently in progress, and thus not included. Auger and radiative damping are taken into account
by means of an optical potential \citep{rob95, gor96, gor00}
where the resonance energy with respect to the threshold acquires
an imaginary component. In the present work, the $N$-electron
targets are represented with all the fine structure levels within
the $n=2$ complex. It is important to mention that the {\sc bprm}
approach does not allow the inclusion of CRE in the 
photoionization calculations; therefore, both the initial and
final states correspond to configurations represented with orthogonal
orbitals. Thus, the wave functions for the target states are those
produced with approximation AS1.

{\sc hullac} \citep[Hebrew University Lawrence Livermore Atomic Code,][]{hullac01} is a
multiconfiguration, relativistic computing package that is based on the
relativistic version of the parametric potential method by \citet{klapisch77},
and employs a factorization-interpolation method within the framework of the
distorted wave approximation \citep{bar-shalom88}.
It includes the Breit interaction for relativistic configuration averages and
can take into account part of the correlation effects by allowing different potentials
for each group of configurations.
Its newest version \citep{hullac09}, which is used here, incorporates a number of improvements
such as important corrections to the photoionization subroutines.
For the present work we calculated only transition energies and direct photoionization cross sections,
but {\sc hullac} can also efficiently compute photo-autoionization resonances by means of the isolated resonance approximation \citep{oreg91}, which are subsequently superimposed on the
continuum photoionization cross section.
Moreover, attempts have been made to adapt {\sc hullac} to calculate the quantum interference of resonances with the continuum that leads to Fano-type asymmetric profiles \citep{behar00, behar04}.
However, these calculations with {\sc hullac} are beyond the scope of the present paper.


\section{Energy levels}

Energies have been computed for both valence and K-vacancy levels in the five
approximations delineated in Section~\ref{numerical_methods} and with {\sc hullac}.
A comparison of approximations AS1 with AS2 provides an estimate of CRE while those of AS1
with AS3 and HF1 with HF2 give measures of out-of-complex
CI. Also, level energy accuracy can be bound with
a comparison of AS2 and HF1, that is, from two physically comparable approximations
but calculated with two independent numerical codes.

\begin{figure}
\epsscale{1.} \plotone{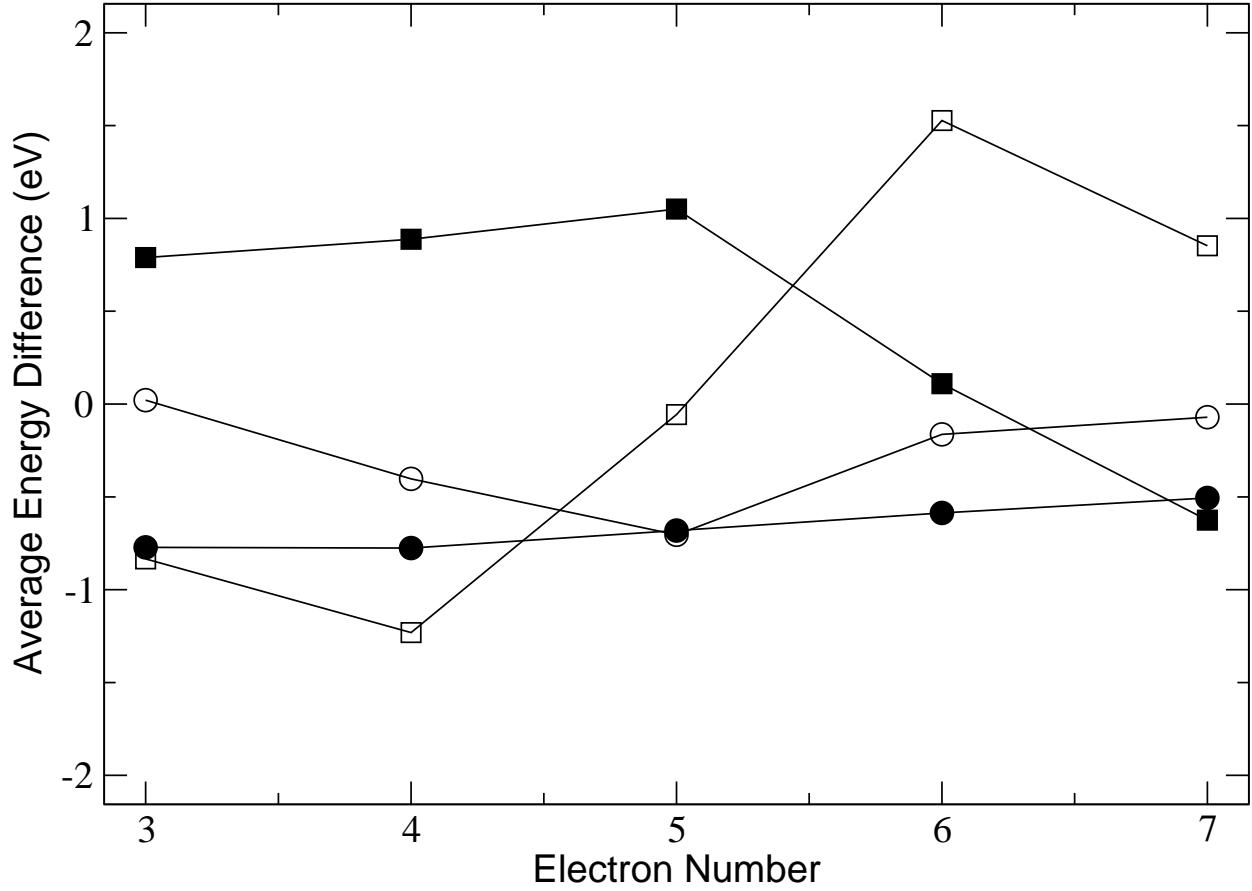}
\caption{Average level energy differences with respect to approximation AS1 for ions of the
nitrogen isonuclear sequence with electron number $3\leq N\leq 7$. Filled circles:
valence level energies computed with AS2. Open circles: valence level energies
computed with AS3. Filled squares: K-vacancy level energies computed with AS2.
Open squares: K-vacancy level energies computed with AS3. \label{AS}}
\end{figure}

In Figure~\ref{AS}, average energy differences for AS1 vs. AS2  and AS1 vs. AS3
are plotted for each ionic species, $3\leq N \leq 7$. It may be seen that while
core relaxation effects lowers the valence-level energies by around
0.5--0.8 eV, it raises by a similar amount those for the K-vacancy levels in species
with electron number $3\leq N \leq 5$; as a consequence, transition wavelengths
for these ions are expected to be shorter due to this effect. Although $n=3$
CI also causes a lowering (less than 0.7 eV) of the valence level energies
(see Figure~\ref{AS}), the impact on the K-vacancy level energies is more pronounced
(as large as 1.5 eV): it mainly lowers levels for species with $N<5$ and raises
those for $N>5$. The latter result is mostly due to K-vacancy levels from
the $n=2$ and $ n=3$ complexes intermixing in the lowly ionized members.

\begin{figure}
\epsscale{1.} \plotone{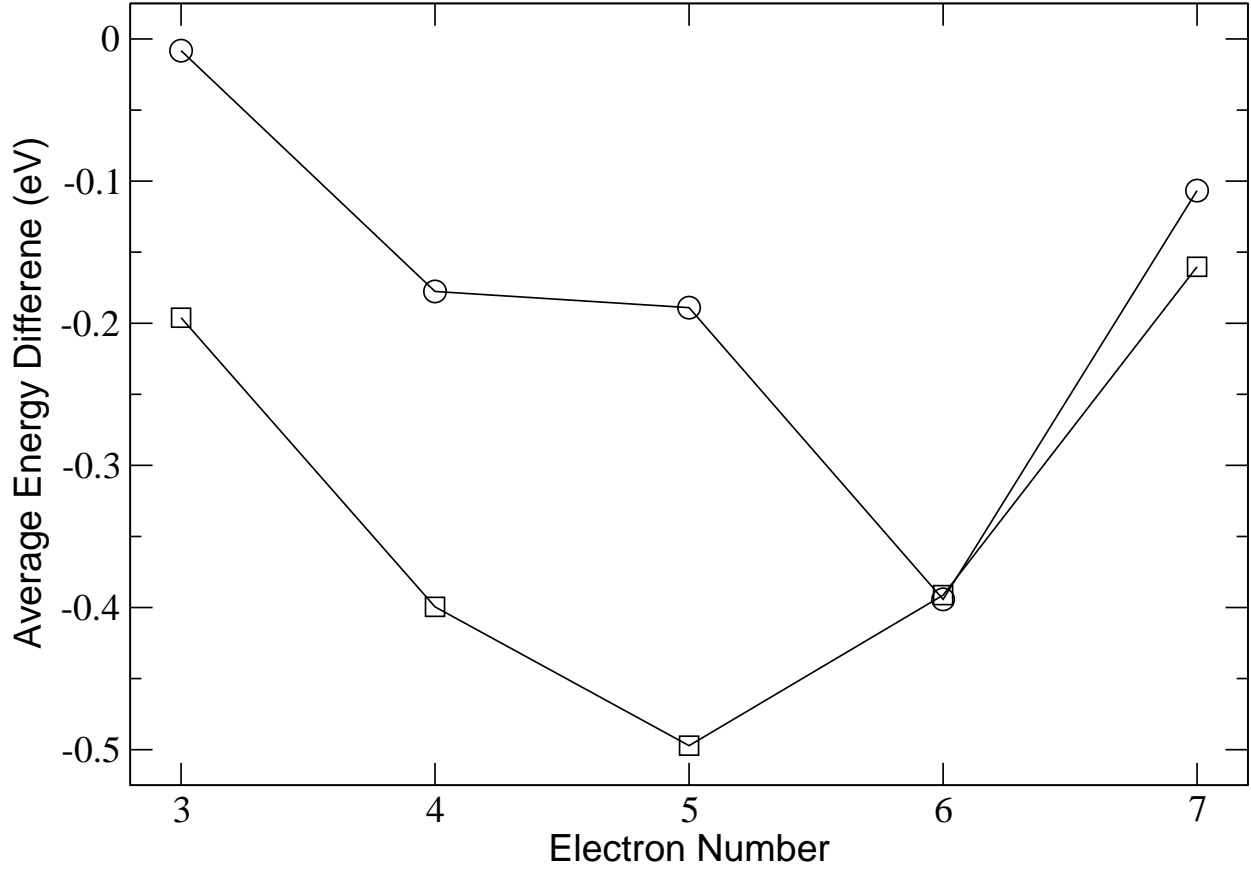}
\caption{Average level energy differences with respect to approximation HF1 for ions of the
nitrogen isonuclear sequence with electron number $3\leq N\leq 7$. Open circles:
valence level energies computed with HF2. Open squares: K-vacancy level energies
computed with HF2. \label{HFR}}
\end{figure}

Figure~\ref{HFR} shows the corresponding quantities for HF1 and HF2.
As shown in Figure~\ref{HFR}, the effect of CI on the energies
computed with {\sc hfr} is somewhat different as they are decreased (less than
0.5 eV) for both valence and K-vacancy levels, the minimum occurring in ions
with $N=5$ and $N=6$. Level energy differences between the AS2 and HF1 data sets
are within 0.5 eV which is a reliable accuracy ranking of the present
level energies.

Computed level energies are compared with the few spectroscopic measurements
available ($2\leq N\leq 3$) in Table~\ref{K_e}. It may be seen that CRE
in {\sc autostructure} (approximation AS2) in general reduce differences with experiment.
Furthermore, {\sc hfr} and {\sc hullac} seem to provide better energies than {\sc autostructure};
differences of {\sc hfr} ({\sc hullac}) with experiment not being larger than 1.53~eV (1.19~eV).
In the He-like system, {\sc hullac} is particularly accurate for triplet states and less so for singlet states.
The accurate approximation of HF2 is compared with previously computed
term energies in Table~\ref{Li_E}. Although we would not quote present term energies
with the same number of significant figures as previous results, HF2 values are consistently
lower.

Fine-structure level splittings in HF2 can be problematic as shown in Table~\ref{split}. 
It may be seen that HF2 gives a value in good accord (1\%) with beam-foil
measurements for the $\Delta E(^5{\rm P}_2, ^5{\rm P}_1)$ splitting of the ${\rm 1s2s2p^2\ ^5P}$
K-vacancy term of N~{\sc iv} but an intolerably discrepant one (factor of 2) for
$\Delta E(^5{\rm P}_3, ^5{\rm P}_2)$. On the other hand, both AS2 splittings are in
reasonable agreement (10\%) with experiment and with results obtained with the relativistic
Raleigh--Ritz variational method \citep{yan95, wan06} and MCDF \citep{hat83, har84}. This
problem has been shown by \citet{hat83} to be due to the neglect of the Breit interaction
which is the case in HF2.

The most stringent accuracy requirements for the energies come from the capabilities of 
the astronomical instruments which can observe these transitions. Current instruments, 
the {\it Chandra} and {\it XMM-Newton} gratings, have a resolving power which is nominally 
$\varepsilon/\Delta\varepsilon \leq$ 1000, which imposes a resolution of $0.4-0.5$ eV
in the energy region of the nitrogen K lines ($400-500$ eV), essentially the same accuracy we
have achieved in the present calculations. However, in spectra with good statistical signal-to-noise 
it is possible to determine line centroids to a factor $\sim$3 more accurately than this.
Future instruments, principally the grating instruments considered for the International X-ray 
Observatory (IXO), may also have resolving power as high as 3000. Knowledge of transition wavelengths 
and energy levels with this precision is needed for truly unambiguous identification of observed 
features and comparison with models. Clearly, precise laboratory measurements are irreplaceable
requisites in the theoretical fine-tuning of these calculations in order to reduce 
the current uncertainties.


\section{Wavelengths}

The accuracy of computed wavelengths must be determined without a comparison with
measurements due to the scarcity of the latter. CRE and $n=3$ CI in {\sc autostructure}
impact wavelengths in an opposite manner to that displayed for the K-vacancy levels
in Figure~\ref{AS}. Specifically, core relaxation
on average shortens wavelengths by as much as 150~m\AA\ in ions with
$2\leq N \leq 5$ while the effect is less pronounced on the higher members. CI increases
wavelengths by $\sim$50~m\AA\ for $N \leq 4$ and decreases them by as much as 150~m\AA\
for $N \geq 5$. CI in {\sc hfr} in general increases wavelengths
with a maximum of 25~m\AA\ at $N=5$.

Wavelengths computed with AS2 are on average $7\pm 21$~m\AA\ greater
than HF1. This finding can be further appreciated in a comparison
with the few available measured wavelengths (see Table~\ref{lambda}) where computed values
are always greater. HF1 differences with experiment are the smallest (less than 37~m\AA)
while CRE in {\sc autostructure} (AS2) also reduces discrepancies.

\begin{figure}
\epsscale{1.} \plotone{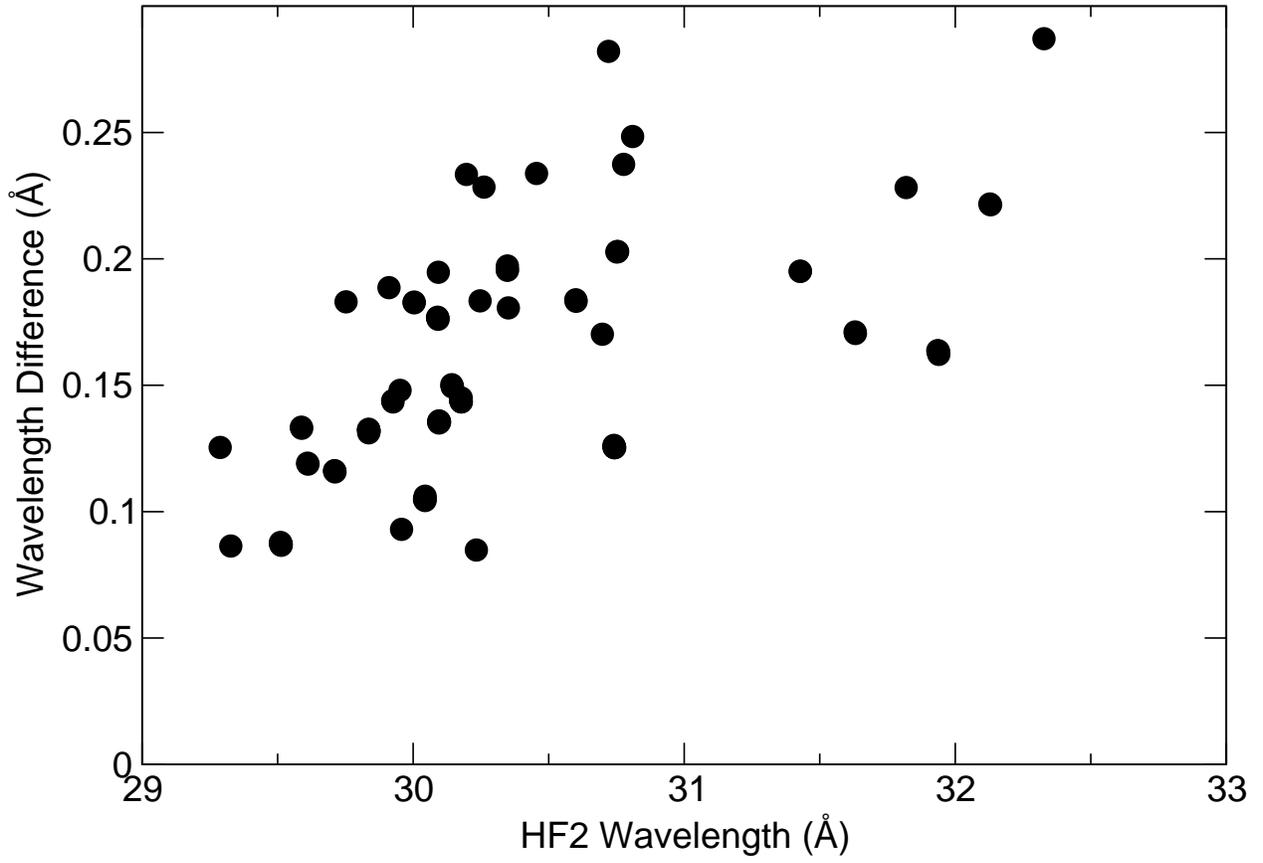}
\caption{Wavelength differences between MCDF \citep{che87} and HF2 for \ion{N}{4}.
An average difference of $155\pm 45$~m\AA\ is observed. \label{mcdf_be}}
\end{figure}

\begin{figure}
\epsscale{1.} \plotone{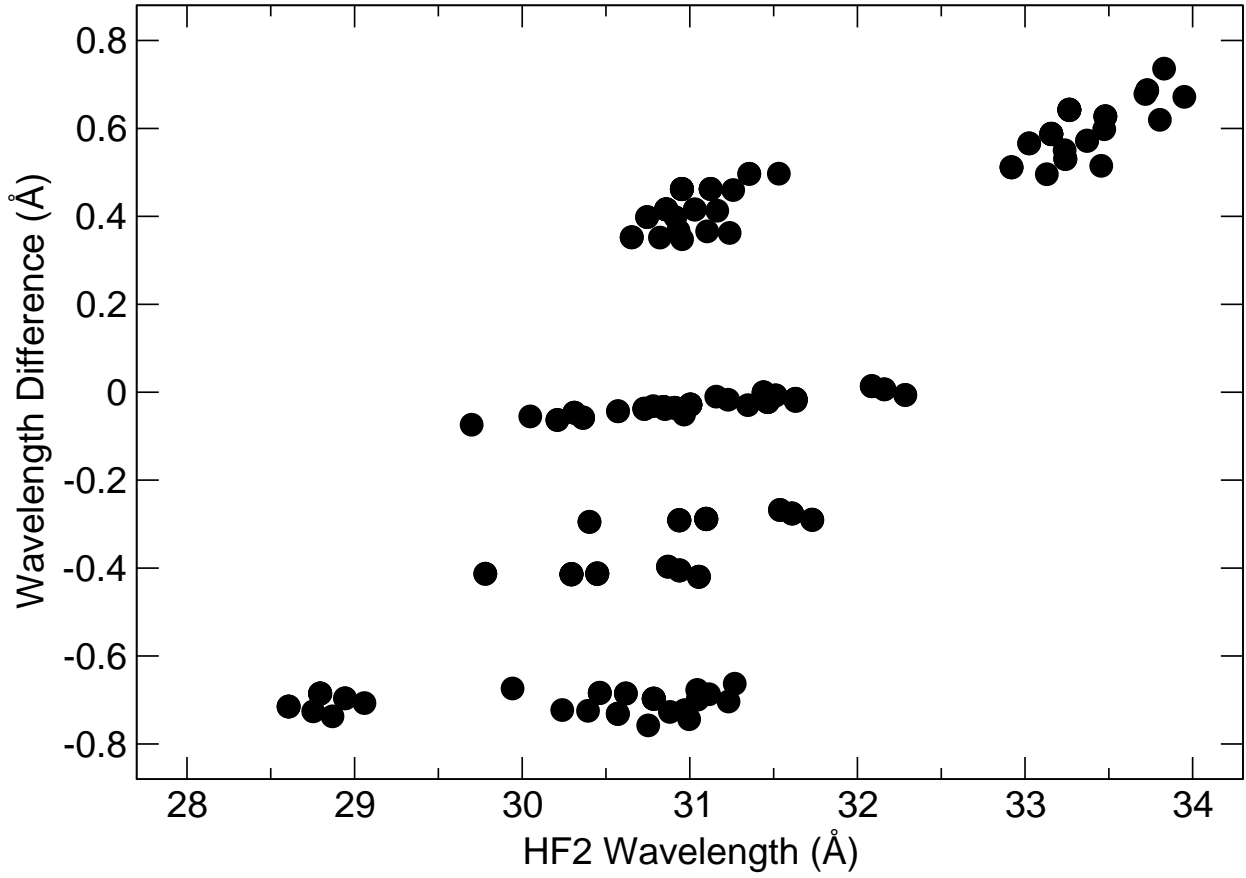}
\caption{Wavelength differences between MCDF \citep{che97} and HF2 for \ion{N}{2}.
Differences as large as 800~m\AA\ are observed. \label{mcdf_c}}
\end{figure}

Wavelengths for K transitions in nitrogen ions have been previously computed with the MCDF
method by \citet{che86, che87, che88, che97}. While reasonable agreement is found with HF2
for species with $N=3$ (average difference of $13\pm 25$~m\AA) and $N=5$ (average difference of
$-4\pm 33$~m\AA), puzzling discrepancies are found for those with $N=4$ and $N=6$.
As shown in Figure~\ref{mcdf_be}, the MCDF wavelengths of \citet{che87} show the large average
difference with respect to HF2 of $155\pm 45$~m\AA; i.e., on average, they are significantly longer. The
inconsistent situation in the C-like ion is somewhat different (see Figure~\ref{mcdf_c})
where the average difference with HF2 is now $-45\pm 449$~m\AA\ showing a very wide and clustered
scatter with questionable deviations as large as 800~m\AA. These comparisons
lead us to conclude that the wavelengths computed with HF2, our best approximation, are
accurate to better than 100~m\AA.


\section{$A$-coefficients}

By comparing $A$-coefficients computed with approximations AS1 and AS2, CRE on the K
radiative decay process may be estimated. Discarding transitions subject to cancellation
which always display large differences, it is found that, for $\log A\geq 10$, CRE generally
causes differences not greater than 20\%. However, larger discrepancies (54\%) are found for
transitions undergoing multiple electron jumps such as those tabulated in Table~\ref{A_cre}.
$A$-coefficients for these peculiar transitions computed with HF1, which should be comparable
to AS2, are also included in Table~\ref{A_cre} finding good agreement (within 15\%);
in fact, differences between $A$-coefficients computed with approximations AS2 and HF1 are
in general within 22\%. Furthermore, CI from the $n=3$ complex tends to decrease
$A$-coefficients with $\log A\geq 10$, AS3 being  on average 16\% lower than AS1 and
HF2 6\% lower than HF1.

\begin{figure}
\epsscale{1.} \plotone{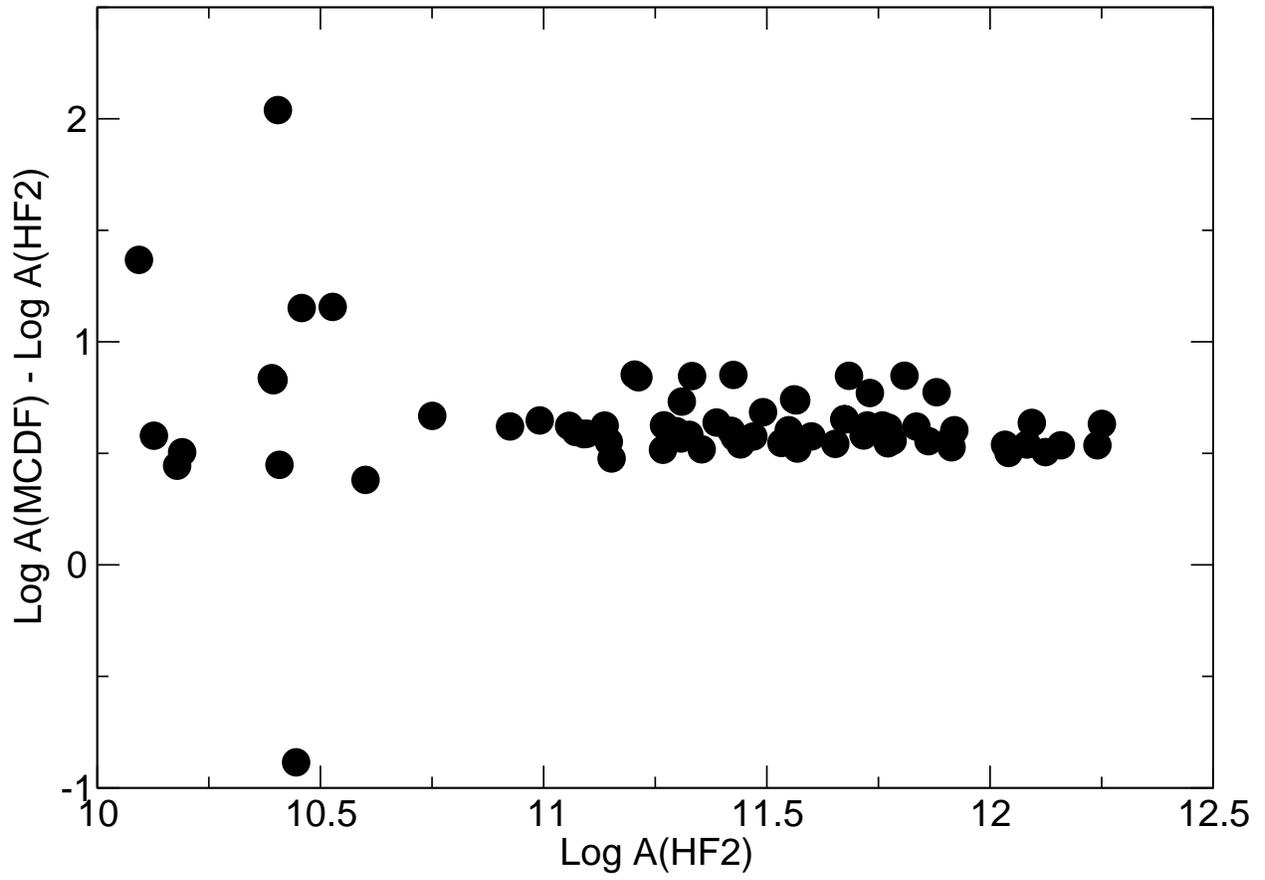}
\caption{$A$-coefficient differences (s$^{-1}$) between MCDF \citep{che97} and HF2 for \ion{N}{2}.
It is found that MCDF is on average higher by a factor of 4. \label{A_mcdf}}
\end{figure}

MCDF $A$-coefficients by \citet{che86, che87, che88, che97} agree with HF2 to around 25\%
except for $N=6$ where they are found to be, on average, higher by a factor of 4
(see Figure~\ref{A_mcdf}). This outcome certainly questions the accuracy of the MCDF
$A$-coefficients by \citet{che97} for C-like nitrogen. We find that for
$\log A\geq 10$ present $A$-coefficients are accurate to within 20\% for transitions
not affected by cancellation.


\section{Radiative widths}

A comparison of AS1 and AS2 radiative widths for the K-vacancy levels indicates that
CRE are mainly noticeable in the highly ionized species, namely those with $N\leq 4$, where on
average the radiative widths are increased by around 10\%. On the other hand, the inclusion of
levels from the $n=3$ complex in the CI expansion (AS3) leads to slightly reduced radiative widths (less
than 7\%) with respect to AS1 for the lowly ionized members ($N\geq 5$). Larger reductions ($\sim$20\%)
are also observed for the lowly ionized species between HF2 (which contains CI from the $n=3$ complex) and
HF1 (which contains CI only from the $n=2$ complex).

A remarkable exception is the radiative width of the ${\rm 1s2s^22p^3\ ^5S^o_2}$ level in the C-like ion.
For this level, AS2 gives $A_j=\sum_i A_{ji}=7.09\times 10^6$~s$^{-1}$
in good agreement with HF1 ($7.65\times 10^6$~s$^{-1}$); however, the AS3 and HF2 radiative widths are
respectively $A_j=5.31\times 10^8$~s$^{-1}$ and $A_j=3.77\times 10^9$~s$^{-1}$, i.e. around two orders
of magnitude larger. The reason for this huge increase when levels from $n=3$ complex are included
in the CI expansion may be appreciated in Table~\ref{quintet}. Within the $n=2$ complex, the
${\rm 1s2s^22p^3\ ^5S^o_2}$ decays radiatively to the ${\rm 1s^22s^22p^2\ ^3P_j}$ ground levels
via two spin-forbidden K$\alpha$ transitions which have small $A$-coefficients ($\lesssim 10^7$~s$^{-1}$). When
the $n=3$ complex is taken into account, $3\rightarrow 2$ spin-allowed channels appear which exhibit
considerably larger $A$-coefficients that add up to the quoted enhancement. The observed discrepancy
between AS3 and HF2 (a factor of 7) are due to severe cancellation in the $\Delta n\ne 0$ transitions.

Radiative widths computed for ions with electron number $3\leq N\leq 6$ with the MCDF method
\citep{che86, che87, che88, che97} agree with HF2 to around 20\% except for the C-like species
where MCDF is a factor of 4.6 higher.


\section{Auger widths}

\begin{figure}
\epsscale{1.} \plotone{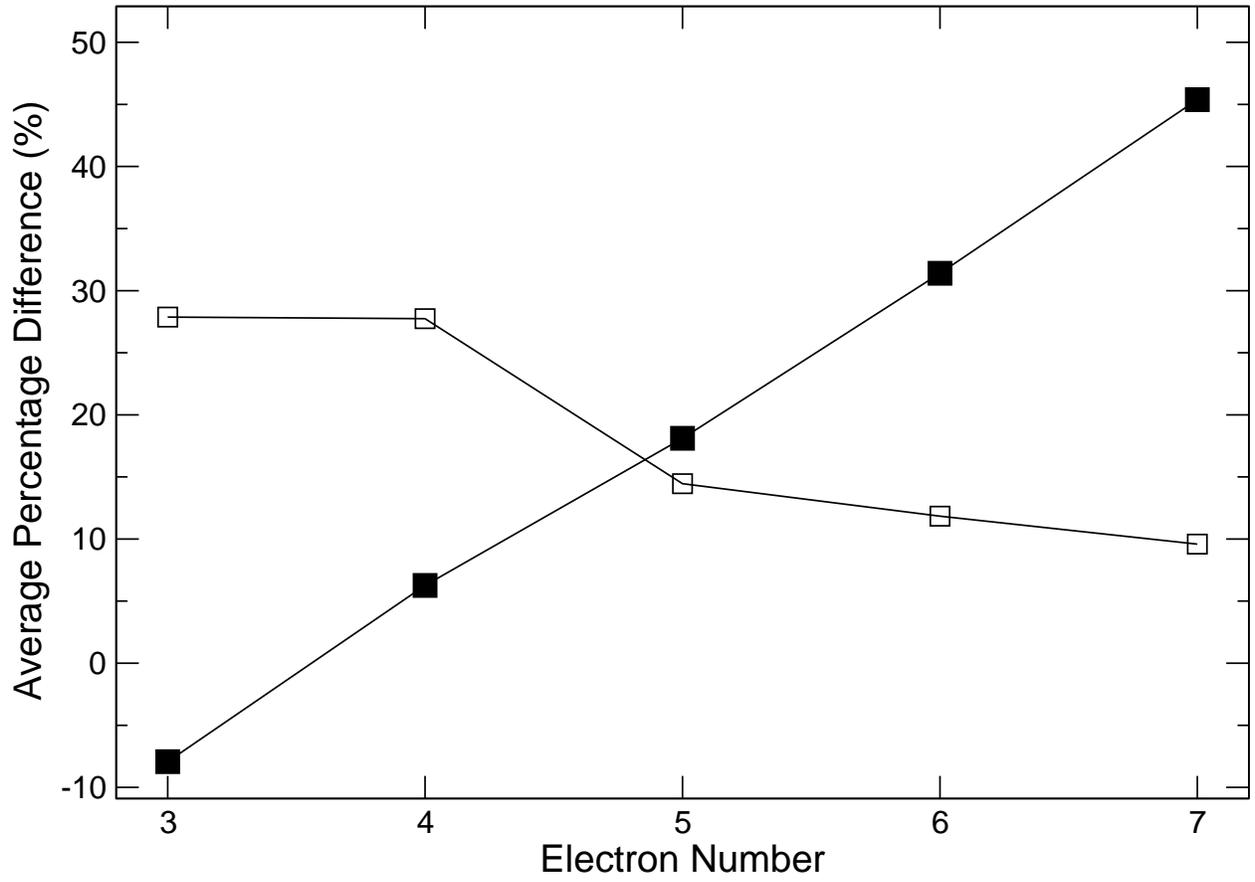}
\caption{Average percentage difference between Auger widths ($\log A_{\rm a}\geq 12$)
computed with the AS2 and AS1 approximations (filled squares) and with AS3 and AS1
(open squares). \label{Au_1}}
\end{figure}

\begin{figure}
\epsscale{1.} \plotone{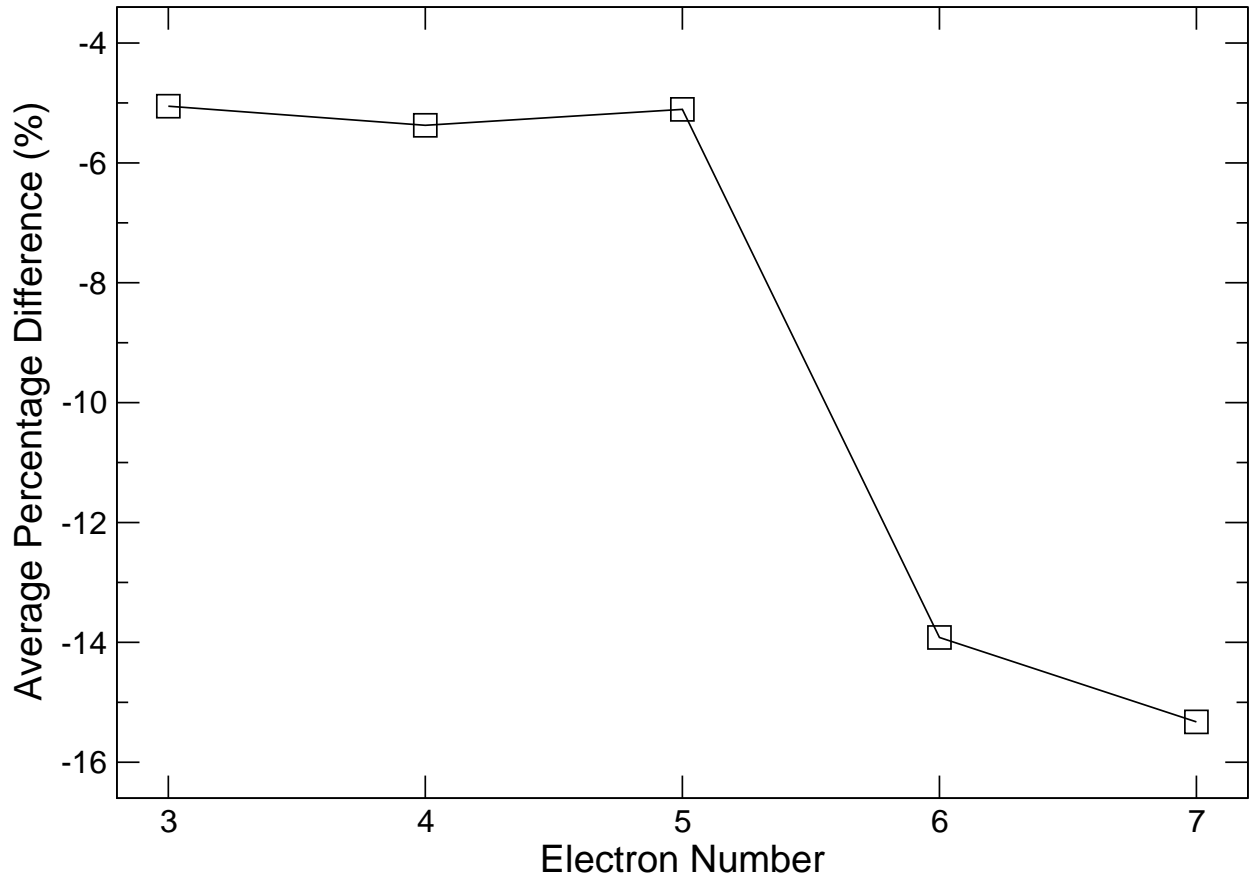}
\caption{Average percentage difference between Auger widths ($\log A_{\rm a}\geq 12$)
computed with the HF2 and HF1. \label{Au_2}}
\end{figure}

Auger widths computed with AS2 and HF1 agree to within 20\% but are sensitive to both
CRE and CI as depicted in Figures~\ref{Au_1}--\ref{Au_2}. A
comparison of AS1 and AS2 shows that, on average, CRE effects increase Auger widths
with $\log A_{\rm a}\geq 12$ linearly as a function of the ion electron number $N$.
In {\sc autostructure} CI also increases Auger widths, particularly for highly
ionized species (around 28\% for ions with $3\leq N\leq 4$) while in {\sc hfr} Auger
widths are decreased (up to 15\% for the lowly ionized members). We believe these contrasting
outcomes are due to the way orbitals have been optimized in {\sc autostructure}.

Excluding the C-like ion, MCDF Auger widths by \citet{che86, che87, che88, che97}
with $\log A_{\rm a}\geq 12$ in general agree with HF2 to around 20\%. Larger
discrepancies are encountered for a handful of K-vacancy levels listed in Table~\ref{Au}
which are mainly caused by level admixture. It may be seen therein that Auger widths
computed with our different approximations also display a wide scatter thus supporting
this diagnostic. For the C-like ion, the MCDF Auger widths are, on average, a factor
of 3 higher than HF2 and thus believed to be of poor quality.

In Table~\ref{Au_li}, Auger widths for the 1s2s2p levels of \ion{N}{5} computed
with the Breit--Pauli saddle-point complex-rotation method \citep{dav89} are compared
with AS2, HF2, and the MCDF values of \citet{che86}. The agreement with AS2 is within 15\%
while very large differences are found for the HF2 ${\rm ^4P^o_j}$ levels which are most surely
due to the neglect of the two-body Breit interaction in {\sc hfr}. The accord with
MCDF is around 25\% except for the ${\rm 1s2s2p\ ^2P^o}$ term where
a discrepancy of a factor of 2 is encountered. The latter is difficult to explain.

Auger widths for K-vacancy terms in \ion{N}{4} computed with the Breit--Pauli saddle-point
complex-rotation method \citep{lin01, lin02, zha05} are compared with AS2, HF2, and MCDF
\citep{che87} in Table~\ref{Au_be}. The level of agreement with AS2 and HF2 is around
20\% except for the values quoted by \citet{zha05} for the ${\rm 1s2p^3\ ^3P^o}$ and
${\rm ^3D^o}$ terms which are discrepant by about 50\%, in contrast with the MCDF Auger widths
for these two terms which agree with AS2 and HF2 to within 10\%. The rest of the MCDF Auger widths
are in good agreement except for the ${\rm 1s2s2p^2\ ^3P}$ term which
has already been singled out in Table~\ref{Au} as being sensitive to level mixing.


\section{Photoabsorption cross sections}

\begin{figure}
\epsscale{1.} \plotone{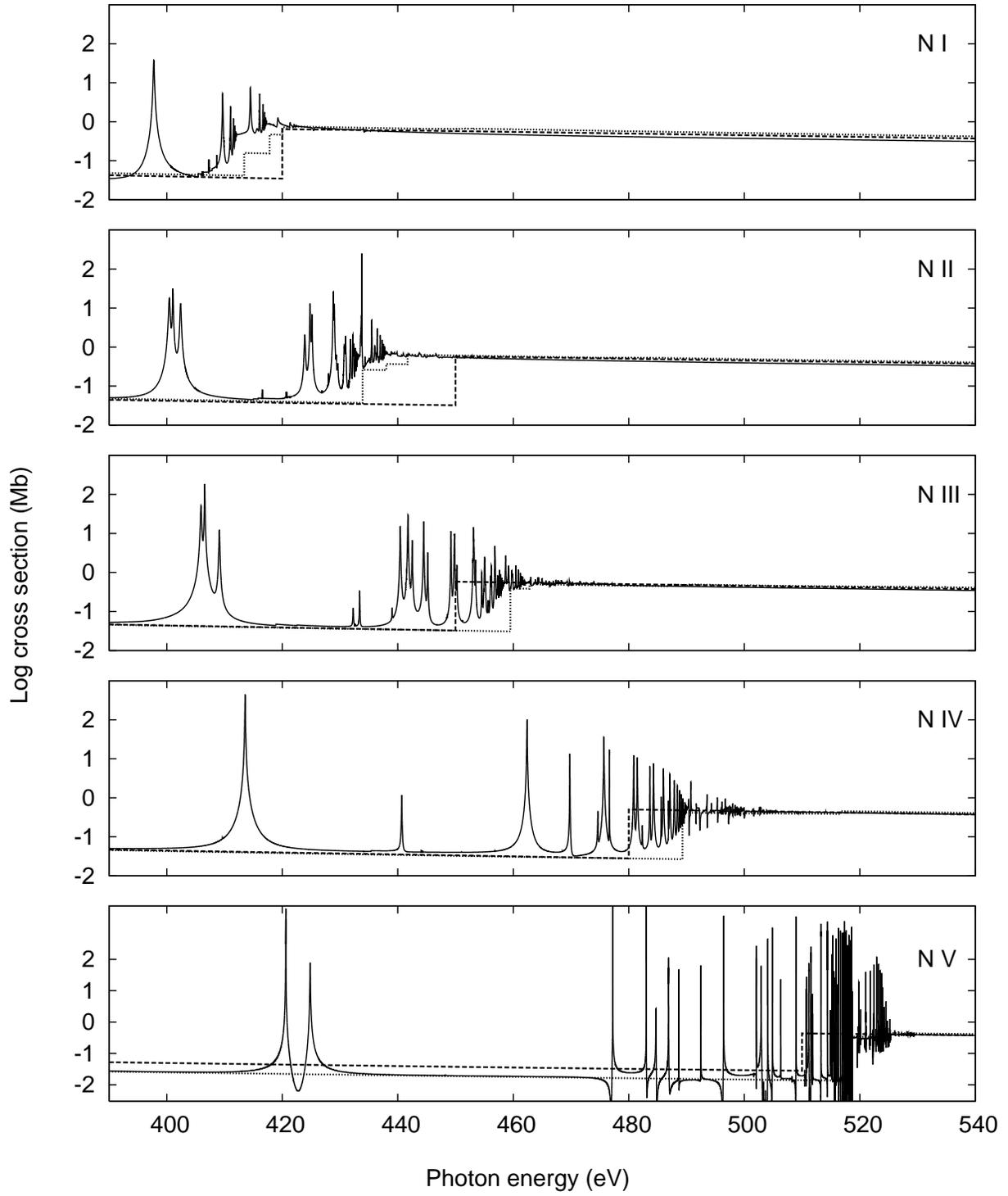}
\caption{High-energy photoabsorption cross sections for nitrogen ions in the K-edge
region. Solid curve: {\sc bprm}. Dotted curve: {\sc hullac}. Dashed curve:
\cite{rei79}.\label{xs}}
\end{figure}

In Figure~\ref{xs}, we show the high-energy photoabsorption cross
sections of \ion{N}{1} -- \ion{N}{5} computed with the {\sc bprm} package.
Intermediate coupling has been used by implementing the AS1 approximation to describe
the target wave functions. In order to resolve accurately the K-threshold
region, radiative and Auger damping are taken into account as described
by \citet{gor00} using the Auger widths calculated with the HF2 approximation.
For comparison, we have included the photoionization cross sections obtained with
the {\sc hullac} code and those by \citet{rei79}, the latter calculated in a central-field
potential. This comparison shows that the K-threshold energies of
{\sc bprm} and {\sc hullac} are in very good accord (within 1 eV) and 
the background cross sections to within $\sim 10\%$.
Note that for the sake of comparison, we have used {\sc hullac} only to compute
the direct bound-free photoionization cross section and not the resonances.
Background cross sections by \cite{rei79}
are in excellent agreement with {\sc bprm} for all ions, but K-edge positions and
structures are clearly discrepant. The present {\sc bprm} calculations result in
smeared K edges due to the dominance of the Auger spectator (KLL) channels over
the participator (KL$n$) channels. This overall behavior is similar to that
reported in previous calculations \citep{kal04, wit09}, and in particular to that
displayed by the corresponding oxygen ions \citep[see Figure~5 in][]{gar05}.

\section{Machine-readable tables}

We include two machine readable tables. For nitrogen ions with electron occupancy
$N=1-7$ and for both valence and Auger levels, Table~\ref{elec1} tabulates the 
spin multiplicity, total orbital angular momentum and total angular momentum quantum
numbers, configuration assignment, energy, and radiative and Auger widths. Fields not
applicable or not computed are labeled with the identifier ``$-9.99$E$+2$". For K
transitions, Table~\ref{elec2} tabulates the wavelength, $A$-coefficient, and $gf$-value. The
printed stub versions list the corresponding data only for the Li-like ion ($N=3$).
The photoionization curves may be requested from the corresponding author.

\section{Discussion and conclusions}

Detailed calculations have been carried out on the atomic properties
of K-vacancy states in ions of the nitrogen isonuclear sequence. Data sets
containing energy levels, wavelengths, $A$-coefficients and radiative and Auger
widths for K-vacancy levels have been computed with the atomic
structure codes {\sc hfr} and {\sc autostructure}. High-energy photoionization
and photoabsorption cross sections for members with electron occupancies $N\geq 3$
have been calculated with the {\sc bprm} and {\sc hullac} codes.

Our best approximation (HF2) takes into account both core-relaxation effects
and configuration interaction from the $n=3$ complex. By comparing results from different
approximations with previous theoretical work and the few spectroscopic
measurements available, we conclude that level energies and wavelengths for
all the nitrogen ions considered in the present calculations can be quoted to be
accurate to within 0.5~eV and 100 m\AA, respectively. The accuracy of
$A$-coefficients and radiative and Auger widths is estimated at approximately
20\% for transitions neither affected by cancelation nor strong level admixture.
Outstanding discrepancies are found with some MCDF data, in particular wavelengths
and $A$-coefficients for the C-like ion which are believed to be due to numerical
error by \citet{che97}.

We have also presented detailed photoabsorption cross sections of nitrogen
ions in the near K-threshold region. Due to the lack of previous experimental
and theoretical results, we have also performed simpler calculations using the
{\sc hullac} code in order to check for consistency and to estimate accuracy.
Comparison of {\sc bprm} and {\sc hullac} indicates that present K-threshold
energies are accurate to within 1~eV. However, background cross sections are in
better agreement with those computed by \citet{rei79} in a central-field
potential for all ions except the Li-like system. These photoabsorption
cross sections and their structures are similar to those displayed by ions
in the same isoelectronic sequence \citep{wit09}, especially to the corresponding
oxygen ions \citep{gar05}.

The present atomic data sets are available on request and will be incorporated in the {\sc xstar} modeling code
in order to generate improved opacities in the nitrogen K-edge region, which will
lead to useful astrophysical diagnostics such as those mentioned in \S~1.

\acknowledgments

We would like to thank the anonymous referee for suggestions that improved the 
clarity of this paper.
This work was funded in part by the NASA Astronomy and Physics Research
and Analysis Program. PP and PQ are research associates of the Belgian
FRS-FNRS. EB acknowledges funding from NASA grant 08-ADP08-0076.
EB and MK thank Michel Busquet for his significant contributions to the revised version
of {\sc hullac} used for this work. This research has made use of NASA's Astrophysics Data System.

\bibliographystyle{apj}
\bibliography{references}

\begin{thebibliography}{59}
\expandafter\ifx\csname natexlab\endcsname\relax\def\natexlab#1{#1}\fi

\bibitem[{{Arav} {et~al.}(2007){Arav}, {Gabel}, {Korista}, {Kaastra}, {Kriss},
  {Behar}, {Costantini}, {Gaskell}, {Laor}, {Kodituwakku}, {Proga}, {Sako},
  {Scott}, \& {Steenbrugge}}]{arav07}
{Arav}, N., {et~al.} 2007, \apj, 658, 829

\bibitem[{{Badnell}(1986)}]{bad86}
{Badnell}, N.~R. 1986, J. Phys. B, 19, 3827

\bibitem[{{Badnell}(1997)}]{bad97}
---. 1997, J. Phys. B, 30, 1

\bibitem[{{Bar-Shalom} {et~al.}(1988){Bar-Shalom}, {Klapisch}, \&
  {Oreg}}]{bar-shalom88}
{Bar-Shalom}, A., {Klapisch}, M., \& {Oreg}, J. 1988, \pra, 38, 1773

\bibitem[{{Bar-Shalom} {et~al.}(2001){Bar-Shalom}, {Klapisch}, \&
  {Oreg}}]{hullac01}
---. 2001, J. Quant. Spectrosc. Radiat. Transfer, 71, 169

\bibitem[{{Bautista} {et~al.}(2003){Bautista}, {Mendoza}, {Kallman}, \&
  {Palmeri}}]{bau03}
{Bautista}, M.~A., {Mendoza}, C., {Kallman}, T.~R., \& {Palmeri}, P. 2003,
  \aap, 403, 339

\bibitem[{{Behar} {et~al.}(2000){Behar}, {Jacobs}, {Oreg}, {Bar-Shalom}, \&
  {Haan}}]{behar00}
{Behar}, E., {Jacobs}, V.~L., {Oreg}, J., {Bar-Shalom}, A., \& {Haan}, S.~L.
  2000, \pra, 62, 030501

\bibitem[{{Behar} {et~al.}(2004){Behar}, {Jacobs}, {Oreg}, {Bar-Shalom}, \&
  {Haan}}]{behar04}
---. 2004, \pra, 69, 022704

\bibitem[{{Behar} \& {Netzer}(2002)}]{behar02}
{Behar}, E., \& {Netzer}, H. 2002, \apj, 570, 165

\bibitem[{{Behar} {et~al.}(2003){Behar}, {Rasmussen}, {Blustin}, {Sako},
  {Kahn}, {Kaastra}, {Branduardi-Raymont}, \& {Steenbrugge}}]{behar03}
{Behar}, E., {Rasmussen}, A.~P., {Blustin}, A.~J., {Sako}, M., {Kahn}, S.~M.,
  {Kaastra}, J.~S., {Branduardi-Raymont}, G., \& {Steenbrugge}, K.~C. 2003,
  \apj, 598, 232

\bibitem[{{Behar} {et~al.}(2001){Behar}, {Sako}, \& {Kahn}}]{behar01}
{Behar}, E., {Sako}, M., \& {Kahn}, S.~M. 2001, \apj, 563, 497

\bibitem[{{Beiersdorfer} {et~al.}(1999){Beiersdorfer}, {L{\'o}pez-Urrutia},
  {Springer}, {Utter}, \& {Wong}}]{bei99}
{Beiersdorfer}, P., {L{\'o}pez-Urrutia}, J.~R.~C., {Springer}, P., {Utter},
  S.~B., \& {Wong}, K.~L. 1999, Rev. Sci. Instrum., 70, 276

\bibitem[{{Berrington} {et~al.}(1987){Berrington}, {Burke}, {Butler}, {Seaton},
  {Storey}, {Taylor}, \& {Yan}}]{ber87}
{Berrington}, K.~A., {Burke}, P.~G., {Butler}, K., {Seaton}, M.~J., {Storey},
  P.~J., {Taylor}, K.~T., \& {Yan}, Y. 1987, J. Phys. B, 20, 6379

\bibitem[{{Berry} {et~al.}(1982){Berry}, {Brooks}, {Cheng}, {Hardis}, \&
  {Ray}}]{ber82}
{Berry}, H.~G., {Brooks}, R.~L., {Cheng}, K.~T., {Hardis}, J.~E., \& {Ray}, W.
  1982, \physscr, 25, 391

\bibitem[{{Brinkman} {et~al.}(2002){Brinkman}, {Kaastra}, {van der Meer},
  {Kinkhabwala}, {Behar}, {Kahn}, {Paerels}, \& {Sako}}]{brinkman02}
{Brinkman}, A.~C., {Kaastra}, J.~S., {van der Meer}, R.~L.~J., {Kinkhabwala},
  A., {Behar}, E., {Kahn}, S.~M., {Paerels}, F.~B.~S., \& {Sako}, M. 2002,
  \aap, 396, 761

\bibitem[{{Chen}(1986)}]{che86}
{Chen}, M.~H. 1986, At. Data Nucl. Data Tables, 34, 301

\bibitem[{{Chen} \& {Crasemann}(1987)}]{che87}
{Chen}, M.~H., \& {Crasemann}, B. 1987, At. Data Nucl. Data Tables, 37, 419

\bibitem[{{Chen} \& {Crasemann}(1988)}]{che88}
---. 1988, At. Data Nucl. Data Tables, 38, 381

\bibitem[{{Chen} {et~al.}(1997){Chen}, {Reed}, {McWilliams}, {Guo}, {Barlow},
  {Lee}, \& {Walker}}]{che97}
{Chen}, M.~H., {Reed}, K.~J., {McWilliams}, D.~M., {Guo}, D.~S., {Barlow}, L.,
  {Lee}, M., \& {Walker}, V. 1997, At. Data Nucl. Data Tables, 65, 289

\bibitem[{{Chung}(1990)}]{chu90}
{Chung}, K.~T. 1990, \pra, 42, 645

\bibitem[{{Cowan}(1981)}]{cow81}
{Cowan}, R.~D. 1981, {The Theory of Atomic Structure and Spectra (Berkeley, CA:
  Univ. California Press)}

\bibitem[{{Davis} \& {Chung}(1989)}]{dav89}
{Davis}, B.~F., \& {Chung}, K.~T. 1989, \pra, 39, 3942

\bibitem[{{Eissner} {et~al.}(1974){Eissner}, {Jones}, \& {Nussbaumer}}]{eis74}
{Eissner}, W., {Jones}, M., \& {Nussbaumer}, H. 1974, Comput. Phys. Commun., 8,
  270

\bibitem[{{Favata} {et~al.}(2009){Favata}, {Neiner}, {Testa}, {Hussain}, \&
  {Sanz-Forcada}}]{fav09}
{Favata}, F., {Neiner}, C., {Testa}, P., {Hussain}, G., \& {Sanz-Forcada}, J.
  2009, \aap, 495, 217

\bibitem[{{Garc{\'{\i}}a} {et~al.}(2005){Garc{\'{\i}}a}, {Mendoza}, {Bautista},
  {Gorczyca}, {Kallman}, \& {Palmeri}}]{gar05}
{Garc{\'{\i}}a}, J., {Mendoza}, C., {Bautista}, M.~A., {Gorczyca}, T.~W.,
  {Kallman}, T.~R., \& {Palmeri}, P. 2005, \apjs, 158, 68

\bibitem[{{Gorczyca} \& {Badnell}(1996)}]{gor96}
{Gorczyca}, T.~W., \& {Badnell}, N.~R. 1996, J. Phys. B, 29, L283

\bibitem[{{Gorczyca} \& {McLaughlin}(2000)}]{gor00}
{Gorczyca}, T.~W., \& {McLaughlin}, B.~M. 2000, J. Phys. B, 33, L859

\bibitem[{{Hardis} {et~al.}(1984){Hardis}, {Berry}, {Curtis}, \&
  {Livingston}}]{har84}
{Hardis}, J.~E., {Berry}, H.~G., {Curtis}, L.~J., \& {Livingston}, A.~E. 1984,
  \physscr, 30, 189

\bibitem[{{Hata} \& {Grant}(1983)}]{hat83}
{Hata}, J., \& {Grant}, I.~P. 1983, J. Phys. B, 16, L125

\bibitem[{{Hsu} {et~al.}(1991){Hsu}, {Chung}, \& {Huang}}]{hsu91}
{Hsu}, J.-J., {Chung}, K.~T., \& {Huang}, K.-N. 1991, \pra, 44, 5485

\bibitem[{{Kaastra} {et~al.}(2009){Kaastra}, {de Vries}, {Costantini}, \& {den
  Herder}}]{kaastra09}
{Kaastra}, J.~S., {de Vries}, C.~P., {Costantini}, E., \& {den Herder},
  J.~W.~A. 2009, \aap, 497, 291

\bibitem[{{Kallman} \& {Bautista}(2001)}]{kal01}
{Kallman}, T., \& {Bautista}, M. 2001, \apjs, 133, 221

\bibitem[{{Kallman} {et~al.}(2004){Kallman}, {Palmeri}, {Bautista}, {Mendoza},
  \& {Krolik}}]{kal04}
{Kallman}, T.~R., {Palmeri}, P., {Bautista}, M.~A., {Mendoza}, C., \& {Krolik},
  J.~H. 2004, \apjs, 155, 675

\bibitem[{{Klapisch} \& {Busquet}(2009)}]{hullac09}
{Klapisch}, M., \& {Busquet}, M. 2009, High Ener. Dens. Phys., 5, 105

\bibitem[{{Klapisch} {et~al.}(1977){Klapisch}, {Schwob}, {Fraenkel}, \&
  {Oreg}}]{klapisch77}
{Klapisch}, M., {Schwob}, J.~L., {Fraenkel}, B.~S., \& {Oreg}, J. 1977, J. Opt.
  Soc. Am., 67, 148

\bibitem[{{Leutenegger} {et~al.}(2003){Leutenegger}, {Kahn}, \&
  {Ramsay}}]{leu03}
{Leutenegger}, M.~A., {Kahn}, S.~M., \& {Ramsay}, G. 2003, \apj, 585, 1015

\bibitem[{{Lin} {et~al.}(2002){Lin}, {Hsue}, \& {Chung}}]{lin02}
{Lin}, H., {Hsue}, C.-S., \& {Chung}, K.~T. 2002, \pra, 65, 032706

\bibitem[{{Lin} {et~al.}(2001){Lin}, {Hsue}, \& {Chung}}]{lin01}
{Lin}, S.-H., {Hsue}, C.-S., \& {Chung}, K.~T. 2001, \pra, 64, 012709

\bibitem[{{Miyata} {et~al.}(2008){Miyata}, {Masai}, \& {Hughes}}]{miy08}
{Miyata}, E., {Masai}, K., \& {Hughes}, J.~P. 2008, \pasj, 60, 521

\bibitem[{{Ness} {et~al.}(2003){Ness}, {Starrfield}, {Burwitz}, {Wichmann},
  {Hauschildt}, {Drake}, {Wagner}, {Bond}, {Krautter}, {Orio}, {Hernanz},
  {Gehrz}, {Woodward}, {Butt}, {Mukai}, {Balman}, \& {Truran}}]{ness03}
{Ness}, J.-U., {et~al.} 2003, \apjl, 594, L127

\bibitem[{{Oreg} {et~al.}(1991){Oreg}, {Goldstein}, {Klapisch}, \&
  {Bar-Shalom}}]{oreg91}
{Oreg}, J., {Goldstein}, W.~H., {Klapisch}, M., \& {Bar-Shalom}, A. 1991, \pra,
  44, 1750

\bibitem[{{Palmeri} {et~al.}(2008{\natexlab{a}}){Palmeri}, {Quinet}, {Mendoza},
  {Bautista}, {Garc{\'{\i}}a}, \& {Kallman}}]{pal08a}
{Palmeri}, P., {Quinet}, P., {Mendoza}, C., {Bautista}, M.~A., {Garc{\'{\i}}a},
  J., \& {Kallman}, T.~R. 2008{\natexlab{a}}, \apjs, 177, 408

\bibitem[{{Palmeri} {et~al.}(2008{\natexlab{b}}){Palmeri}, {Quinet}, {Mendoza},
  {Bautista}, {Garc{\'{\i}}a}, {Witthoeft}, \& {Kallman}}]{pal08b}
{Palmeri}, P., {Quinet}, P., {Mendoza}, C., {Bautista}, M.~A., {Garc{\'{\i}}a},
  J., {Witthoeft}, M.~C., \& {Kallman}, T.~R. 2008{\natexlab{b}}, \apjs, 179,
  542

\bibitem[{{Pradhan}(2000)}]{pradhan00}
{Pradhan}, A.~K. 2000, \apjl, 545, L165

\bibitem[{{Ralchenko} {et~al.}(2008){Ralchenko}, {Kramida}, {Reader}, \& {NIST
  ADS Team}}]{ral08}
{Ralchenko}, Y., {Kramida}, A.~E., {Reader}, J., \& {NIST ADS Team}. 2008, NIST
  Atomic Spectra Database, version 3.1.5 (Gaithersburg: NIST),
  http://physics.nist.gov/asd3

\bibitem[{{Ram{\'{\i}}rez} {et~al.}(2008){Ram{\'{\i}}rez}, {Komossa},
  {Burwitz}, \& {Mathur}}]{ram08}
{Ram{\'{\i}}rez}, J.~M., {Komossa}, S., {Burwitz}, V., \& {Mathur}, S. 2008,
  \apj, 681, 965

\bibitem[{{Reilman} \& {Manson}(1979)}]{rei79}
{Reilman}, R.~F., \& {Manson}, S.~T. 1979, \apjs, 40, 815

\bibitem[{{Robicheaux} {et~al.}(1995){Robicheaux}, {Gorczyca}, {Pindzola}, \&
  {Badnell}}]{rob95}
{Robicheaux}, F., {Gorczyca}, T.~W., {Pindzola}, M.~S., \& {Badnell}, N.~R.
  1995, \pra, 52, 1319

\bibitem[{{Sako} {et~al.}(2001){Sako}, {Kahn}, {Behar}, {Kaastra}, {Brinkman},
  {Boller}, {Puchnarewicz}, {Starling}, {Liedahl}, {Clavel}, \&
  {Santos-Lleo}}]{sako01}
{Sako}, M., {et~al.} 2001, \aap, 365, L168

\bibitem[{{Scott} \& {Burke}(1980)}]{sco80}
{Scott}, N.~S., \& {Burke}, P.~G. 1980, J. Phys. B, 13, 4299

\bibitem[{{Scott} \& {Taylor}(1982)}]{sco82}
{Scott}, N.~S., \& {Taylor}, K.~T. 1982, Comput. Phys. Commun., 25, 347

\bibitem[{{Seaton}(1987)}]{sea87}
{Seaton}, M. 1987, J. Phys. B At. Mol. Phys., 20, 6363

\bibitem[{{Shiu} {et~al.}(2001){Shiu}, {Hsue}, \& {Chung}}]{shi01}
{Shiu}, W.~C., {Hsue}, C.-S., \& {Chung}, K.~T. 2001, \pra, 64, 022714

\bibitem[{{Steenbrugge} {et~al.}(2005){Steenbrugge}, {Kaastra}, {Crenshaw},
  {Kraemer}, {Arav}, {George}, {Liedahl}, {van der Meer}, {Paerels}, {Turner},
  \& {Yaqoob}}]{steenbrugge05}
{Steenbrugge}, K.~C., {et~al.} 2005, \aap, 434, 569

\bibitem[{{Wang} \& {Gou}(2006)}]{wan06}
{Wang}, F., \& {Gou}, B.~C. 2006, At. Data Nucl. Data Tables, 92, 176

\bibitem[{{Witthoeft} {et~al.}(2009){Witthoeft}, {Bautista}, {Mendoza},
  {Kallman}, {Palmeri}, \& {Quinet}}]{wit09}
{Witthoeft}, M.~C., {Bautista}, M.~A., {Mendoza}, C., {Kallman}, T.~R.,
  {Palmeri}, P., \& {Quinet}, P. 2009, \apjs, 182, 127

\bibitem[{{Yang} \& {Chung}(1995)}]{yan95}
{Yang}, H.~Y., \& {Chung}, K.~T. 1995, \pra, 51, 3621

\bibitem[{{Yao} {et~al.}(2009){Yao}, {Schulz}, {Gu}, {Nowak}, \&
  {Canizares}}]{yao09}
{Yao}, Y., {Schulz}, N.~S., {Gu}, M.~F., {Nowak}, M.~A., \& {Canizares}, C.~R.
  2009, \apj, 696, 1418

\bibitem[{{Zhang} {et~al.}(2005){Zhang}, {Gou}, \& {Cui}}]{zha05}
{Zhang}, M., {Gou}, B.~C., \& {Cui}, L.~L. 2005, J. Phys. B, 38, 3567

\end{thebibliography}


\clearpage

\clearpage

\clearpage

\clearpage

\clearpage

\clearpage

\clearpage

\clearpage

\begin{deluxetable}{llrrrrrrr}
\tablecaption{Measured and computed $K$-vacancy level energies (eV)
\label{K_e}}
\tablewidth{0pt}
\tablehead{
\colhead{$N$} & \colhead{Level} & \colhead{$E$} & \multicolumn{6}{c}{$E({\rm Th})-E({\rm Expt})$}\\
\cline{4-9}\\
\colhead{} & \colhead{} & \colhead{Expt} & \colhead{AS1} & \colhead{AS2} &
\colhead{AS3} & \colhead{HF1} & \colhead{HF2} & {\sc hullac}\\
}
\startdata
2 & ${\rm 1s2s\ ^3S_1}$                    & 419.80 & $-$1.76 & $-$1.34 & $-$2.35 & $-$0.59 & &0.01\\
2 & ${\rm 1s2p\ ^3P^o_0}$                  & 426.30 & $-$1.85 & $-$0.93 & $-$2.36 & $-$0.56 & &0.09\\
2 & ${\rm 1s2p\ ^3P^o_1}$                  & 426.30 & $-$1.85 & $-$0.93 & $-$2.36 & $-$0.54 & &0.11\\
2 & ${\rm 1s2p\ ^3P^o_2}$                  & 426.33 & $-$1.86 & $-$0.94 & $-$2.37 & $-$0.53 & &0.13\\
2 & ${\rm 1s2s\ ^1S_0}$                    & 426.42 & $-$1.61 & $-$0.21 & $-$2.15 & 0.03    & &0.69\\
2 & ${\rm 1s2p\ ^1P^o_1}$                  & 430.70 & $-$1.64 & $-$0.85 & $-$2.24 & $-$0.24 & &0.67\\
3 & ${\rm 1s(^2S)2s2p(^3P^o)\ ^4P^o_{5/2}}$ & 414.61 & $-$2.49 & $-$1.97 & $-$3.64 & $-$1.48 & $-$1.53 &0.42\\
3 & ${\rm 1s(^2S)2s2p(^3P^o)\ ^2P^o_{3/2}}$ & 421.52\tablenotemark{a} & $-$1.72 & $-$0.80 & $-$2.18 & $-$0.51 & $-$0.62 &0.83\\
3 & ${\rm 1s(^2S)2p^2(^3P)\ ^4P_{5/2}}$      & 425.70 & $-$2.54 & $-$1.64 & $-$3.37 & $-$1.31 & $-$1.33 &1.19\\
\enddata
\tablecomments{Experimental level energies (relative to the ion ground state) from the NIST
      database \citep{ral08}.}
\tablenotetext{a}{Derived from the wavelength measurement of \citet{bei99}.}
\end{deluxetable}


\begin{deluxetable}{llll}
\tabletypesize{\scriptsize}
\tablecaption{Computed total energies (au) for $K$-vacancy terms
\label{Li_E}}
\tablewidth{0pt}
\tablehead{
\colhead{$N$} & \colhead{Term} & \colhead{HF2} & \colhead{Other theory} \\
}
\startdata
3 & ${\rm 1s2s2p\ ^4P^o}$             & $-$33.240 & $-$33.192008\tablenotemark{a}, $-$33.192204\tablenotemark{b} \\
3 & ${\rm 1s(^2S)2s2p(^3P^o)\ ^2P^o}$ & $-$32.952 & $-$32.919222\tablenotemark{a} \\
3 & ${\rm 1s(^2S)2s2p(^1P^o)\ ^2P^o}$ & $-$32.777 & $-$32.768550\tablenotemark{a} \\
4 & ${\rm 1s2s^22p\ ^3P^o}$           & $-$36.219 & $-$36.171232\tablenotemark{c}, $-$36.173064\tablenotemark{d} \\
4 & ${\rm 1s2s2p^2\ ^3S}$             & $-$35.646 & $-$35.615357\tablenotemark{d} \\
4 & ${\rm 1s2s2p^2\ ^3P}$             & $-$35.842 & $-$35.788866\tablenotemark{d} \\
4 & ${\rm 1s2s2p^2\ ^3P}$             & $-$35.542 & $-$35.536868\tablenotemark{d} \\
4 & ${\rm 1s2s2p^2\ ^3D}$             & $-$35.815 & $-$35.785042\tablenotemark{d} \\
4 & ${\rm 1s2s^22p\ ^1P^o}$           & $-$36.081 & $-$36.036967\tablenotemark{e} \\
4 & ${\rm 1s2p^3\ ^1P^o}$             & $-$35.086 & $-$35.082958\tablenotemark{e} \\
4 & ${\rm 1s2s2p^2\ ^1D}$             & $-$35.594 & $-$35.583448\tablenotemark{f} \\
4 & ${\rm 1s2s2p^2\ ^1P}$             & $-$35.445 & $-$35.435017\tablenotemark{f} \\
4 & ${\rm 1s2s2p^2\ ^1S}$             & $-$35.425 & $-$35.415017\tablenotemark{f} \\
4 & ${\rm 1s2s2p^2\ ^5P}$             & $-$36.160 & $-$36.0934586\tablenotemark{g}, $-$36.0934407\tablenotemark{h} \\
4 & ${\rm 1s2p^3\ ^5S^o}$             & $-$35.596 & $-$35.5414665\tablenotemark{g}, $-$35.5413248\tablenotemark{h} \\
4 & ${\rm 1s2p^3\ ^3P^o}$             & $-$35.214 & $-$35.204065\tablenotemark{i} \\
4 & ${\rm 1s2p^3\ ^3D^o}$             & $-$35.387 & $-$35.366601\tablenotemark{i} \\
\enddata
\tablenotetext{a}{Breit--Pauli saddle-point complex-rotation method \citep{dav89}}
\tablenotetext{b}{Relativistic Raleigh--Ritz variational method \citep{hsu91}}
\tablenotetext{c}{Breit--Pauli saddle-point complex-rotation method \citep{chu90}}
\tablenotetext{d}{Breit--Pauli saddle-point complex-rotation method \citep{lin01}}
\tablenotetext{e}{Breit--Pauli saddle-point complex-rotation method \citep{lin02}}
\tablenotetext{f}{Breit--Pauli saddle-point complex-rotation method \citep{shi01}}
\tablenotetext{g}{Relativistic Raleigh--Ritz variational method \citep{wan06}}
\tablenotetext{h}{Relativistic Raleigh--Ritz variational method \citep{yan95}}
\tablenotetext{i}{Breit--Pauli saddle-point complex-rotation method \citep{zha05}}

\end{deluxetable}


\begin{deluxetable}{lllrl}
\tablecaption{Level splittings (cm$^{-1}$) for the ${\rm 1s2s2p^2\ ^5P}$ K-vacancy term of \ion{N}{4}
\label{split}}
\tablewidth{0pt}
\tablehead{
\colhead{Level splitting} & \colhead{Expt\tablenotemark{a}} & \colhead{HF2} & \colhead{AS2} & \colhead{Other theory} \\
}
\startdata
$\Delta E(^5{\rm P}_2, ^5{\rm P}_1)$ & 127$\pm$ 1    & 126 & 119 & 127.07\tablenotemark{b}, 126.9\tablenotemark{c}, 129.19\tablenotemark{d}, 128.94\tablenotemark{e} \\
$\Delta E(^5{\rm P}_3, ^5{\rm P}_2)$ & 79.5$\pm$ 0.8 & 188 &  70 & 78.49\tablenotemark{b}, 78.45\tablenotemark{c}, 86.72\tablenotemark{d}, 86.58\tablenotemark{e} \\
\enddata
\tablenotetext{a}{Beam-foil measurements by \citet{ber82}}
\tablenotetext{b}{Relativistic Raleigh--Ritz variational method \citep{wan06}}
\tablenotetext{c}{Relativistic Raleigh--Ritz variational method \citep{yan95}}
\tablenotetext{d}{MCDF-EAL calculation of \citet{hat83}}
\tablenotetext{e}{MCDF calculation of \citet{har84}}
\end{deluxetable}


\begin{deluxetable}{llllllllr}
\tablecaption{Experimental and theoretical wavelengths (\AA) for nitrogen ions
\label{lambda}}
\tablewidth{0pt}
\tablehead{
\colhead{$N$} & \colhead{Lower level} & \colhead{Upper level} & \colhead{$\lambda$(Expt\tablenotemark{a})} &
      \multicolumn{5}{c}{$\lambda({\rm Th})-\lambda({\rm Expt})$}\\
\cline{5-9}\\
\colhead{} & \colhead{} & \colhead{} & \colhead{} & \colhead{HF1} & \colhead{AS1} &
\colhead{AS2} & \colhead{AS3} & {\sc hullac}
}
\startdata
2 & ${\rm 1s^2\ ^1S_0}$       & ${\rm 1s2s\ ^3S_1}$         & $29.5321(26)$ &        & 0.1266 & 0.0971 & 0.1683 &0.0012\\
2 & ${\rm 1s^2\ ^1S_0}$       & ${\rm 1s2p\ ^3P^o_1}$       & $29.0835(26)$ & 0.0374 & 0.1274 & 0.0642 & 0.1627& --0.0074\\
2 & ${\rm 1s^2\ ^1S_0}$       & ${\rm 1s2p\ ^1P^o_1}$       & $28.7861(22)$ & 0.0167 & 0.1108 & 0.0576 & 0.1515 &--0.0444\\
3 & ${\rm 1s^22s\ ^2S_{1/2}}$ & ${\rm 1s2s2p\ ^2P^o_{3/2}}$ & $29.4135(37)$ & 0.0353 & 0.1208 & 0.0560 & 0.1550 &--0.0573\\
\enddata
\tablenotetext{a}{Wavelength measurements by \citet{bei99}}
\end{deluxetable}


\begin{deluxetable}{lllrrr}
\tablecaption{Core relaxation effects on K radiative decay
\label{A_cre}}
\tablewidth{0pt}
\tablehead{
\colhead{$N$} & \colhead{$j$} & \colhead{$i$} & \multicolumn{3}{c}{$A_{ji}$ (s$^{-1})$}\\
\cline{4-6}\\
\colhead{} & \colhead{} & \colhead{} & \colhead{AS1} & \colhead{AS2} & \colhead{HF1}
}
\startdata
3 & ${\rm 1s2s^2\ ^2S_{1/2}}$     & ${\rm 1s^22p\ ^2P^o_{1/2}}$   & 3.85E$+$10 & 5.36E$+$10 & 4.63E$+$10 \\
3 & ${\rm 1s2s^2\ ^2S_{1/2}}$     & ${\rm 1s^22p\ ^2P^o_{3/2}}$   & 7.62E$+$10 & 1.06E$+$11 & 9.18E$+$10 \\
4 & ${\rm 1s2s^22p\ ^1P^o_{1}}$   & ${\rm 1s^22p^2\ ^1D_{2}}$     & 9.72E$+$10 & 1.36E$+$11 & 1.24E$+$11 \\
5 & ${\rm 1s2s^22p^2\ ^2P_{1/2}}$ & ${\rm 1s^22p^3\ ^2P^o_{1/2}}$ & 1.17E$+$10 & 1.80E$+$10 & 1.78E$+$10 \\
5 & ${\rm 1s2s^22p^2\ ^2P_{3/2}}$ & ${\rm 1s^22p^3\ ^2P^o_{3/2}}$ & 1.54E$+$10 & 2.36E$+$10 & 2.32E$+$10 \\
\enddata

\end{deluxetable}


\begin{deluxetable}{llrr}
\tablecaption{Radiative decay routes of ${\rm 1s2s^22p^3\ ^5S^o_{2}}$ in \ion{N}{2}
\label{quintet}}
\tablewidth{0pt}
\tablehead{
\colhead{$j$} & \colhead{$i$} & \multicolumn{2}{c}{$A_{ji}$ (s$^{-1})$}\\
\cline{3-4}\\
\colhead{} & \colhead{} & \colhead{AS3} & \colhead{HF2}
}
\startdata
${\rm 1s2s^22p^3\ ^5S^o_{2}}$  & ${\rm 1s^22s^22p^2\ ^3P_{1}}$ & 2.04E$+$6 & 8.57E$+$6 \\
                               & ${\rm 1s^22s^22p^2\ ^3P_{2}}$ & 2.97E$+$6 & 2.53E$+$7 \\
                               & ${\rm 1s^22s2p^23d\ ^5P_{1}}$ & 5.07E$+$7 & 3.43E$+$8 \\
                               & ${\rm 1s^22s2p^23d\ ^5P_{2}}$ & 8.36E$+$7 & 5.70E$+$8 \\
                               & ${\rm 1s^22s2p^23d\ ^5P_{3}}$ & 1.17E$+$8 & 7.97E$+$8 \\
                               & ${\rm 1s^22p^33p\ ^5P_{1}}$   & 7.85E$+$7 & 6.99E$+$8 \\
                               & ${\rm 1s^22p^33p\ ^5P_{2}}$   & 1.31E$+$7 & 1.16E$+$9 \\
                               & ${\rm 1s^22p^33p\ ^5P_{3}}$   & 1.83E$+$8 & 1.63E$+$8 \\
\cline{1-4}\\
$\sum_i A_{ji}$                &                               & 5.31E$+$8 & 3.77E$+$9 \\
\enddata

\end{deluxetable}


\begin{deluxetable}{llrrrrrr}
\tabletypesize{\scriptsize}
\tablecaption{Discrepant Auger rates (s$^{-1}$) \label{Au}}
\tablewidth{0pt}
\tablehead{
\colhead{$N$} & \colhead{Level} & \colhead{AS1} & \colhead{AS2} & \colhead{AS3} & \colhead{HF1} & \colhead{HF2} &
\colhead{MCDF\tablenotemark{a}}
}
\startdata
 3 & ${\rm 1s(^2S)2s2p(^3P^o)\ ^2P^o_{1/2}}$   & 1.51E+13 & 5.81E+12 & 2.15E+13 & 8.61E+12 & 7.63E+12 & 1.41E+13 \\
 3 & ${\rm 1s(^2S)2s2p(^3P^o)\ ^2P^o_{3/2}}$   & 1.43E+13 & 5.27E+12 & 2.02E+13 & 8.29E+12 & 7.30E+12 & 1.35E+13 \\
 4 & ${\rm 1s(^2S)2s2p^2(^4P)\ ^3P_0}$         & 3.09E+13 & 1.99E+13 & 4.41E+13 & 1.86E+13 & 1.77E+13 & 3.73E+13 \\
 4 & ${\rm 1s(^2S)2s2p^2(^4P)\ ^3P_1}$         & 3.08E+13 & 1.98E+13 & 4.40E+13 & 1.86E+13 & 1.76E+13 & 3.68E+13 \\
 4 & ${\rm 1s(^2S)2s2p^2(^4P)\ ^3P_2}$         & 3.04E+13 & 1.95E+13 & 4.36E+13 & 1.85E+13 & 1.75E+13 & 3.55E+13 \\
 4 & ${\rm 1s(^2S)2s2p^2(^2P)\ ^1P_1}$         & 2.26E+13 & 2.88E+13 & 2.62E+13 & 2.46E+13 & 2.34E+13 & 1.43E+14 \\
 4 & ${\rm 1s(^2S)2s2p^2(^2S)\ ^1S_0}$         & 1.34E+14 & 1.33E+14 & 1.77E+14 & 1.28E+14 & 1.19E+14 & 1.77E+13 \\
 5 & ${\rm 1s(^2S)2s2p^3(^5S^o)\ ^4S^o_{3/2}}$ & 3.90E+13 & 2.98E+13 & 4.10E+13 & 2.54E+13 & 2.48E+13 & 3.89E+13 \\
 5 & ${\rm 1s(^2S)2s2p^3(^3S^o)\ ^2S^o_{1/2}}$ & 2.34E+13 & 3.09E+13 & 2.88E+13 & 5.30E+13 & 1.31E+14 & 3.49E+13 \\
 5 & ${\rm 1s(^2S)2s2p^3(^1P^o)\ ^2P^o_{1/2}}$ & 1.37E+14 & 1.67E+14 & 1.56E+14 & 1.27E+14 & 3.48E+13 & 1.52E+14 \\
\enddata
\tablenotetext{a}{MCDF computations by \citet{che86, che87, che88}}
\end{deluxetable}


\begin{deluxetable}{llll}
\tablecaption{Auger energy widths (au) for 1s2s2p levels in \ion{N}{5}
\label{Au_li}}
\tablewidth{0pt}
\tablehead{
\colhead{Level} & \colhead{AS2} & \colhead{HF2} &\colhead{Other theory} \\
}
\startdata
${\rm 1s(^2S)2s2p(^3P^o)\ ^4P^o_{1/2}}$ & 1.46E$-$08  & 3.48E$-10$ & 1.50E$-08$\tablenotemark{a}, 1.532E$-$08\tablenotemark{b} \\
${\rm 1s(^2S)2s2p(^3P^o)\ ^4P^o_{3/2}}$ & 4.53E$-$09  & 8.83E$-10$ & 4.98E$-09$\tablenotemark{a}, 3.952E$-$09\tablenotemark{b} \\
${\rm 1s(^2S)2s2p(^3P^o)\ ^4P^o_{5/2}}$ & 4.26E$-$10  &            & 4.587E$-$10\tablenotemark{b}                              \\
${\rm 1s(^2S)2s2p(^3P^o)\ ^2P^o}$       & 1.32E$-$04  & 1.79E$-04$ & 3.31E$-$04\tablenotemark{a}, 1.54E$-$04\tablenotemark{b} \\
${\rm 1s(^2S)2s2p(^1P^o)\ ^2P^o}$       & 1.68E$-$03  & 1.47E$-03$ & 1.14E$-$03\tablenotemark{a}, 1.53E$-$03\tablenotemark{b} \\
\enddata
\tablenotetext{a}{MCDF calculations by \citep{che86}}
\tablenotetext{b}{Breit--Pauli saddle-point complex-rotation method \citep{dav89}}
\end{deluxetable}


\begin{deluxetable}{llll}
\tablecaption{Auger widths (meV) for $K$-vacancy terms in \ion{N}{4}
\label{Au_be}}
\tablewidth{0pt}
\tablehead{
\colhead{Term} & \colhead{AS2} & \colhead{HF2} &\colhead{Other theory} \\
}
\startdata
${\rm 1s2s^22p\ ^3P^o}$           & 84.1 & 71.8 & 77.4\tablenotemark{a}, 79.0\tablenotemark{b} \\
${\rm 1s(^2S)2s2p^2(^4P)\ ^3P}$   & 12.9 & 11.6 & 23.8\tablenotemark{a}, 10.8\tablenotemark{b} \\
${\rm 1s(^2S)2s2p^2(^2D)\ ^3D}$   & 66.9 & 62.6 & 53.4\tablenotemark{a}, 57.7\tablenotemark{b} \\
${\rm 1s(^2S)2s2p^2(^2S)\ ^3S}$   & 31.2 & 29.7 & 26.7\tablenotemark{a}, 28.6\tablenotemark{b} \\
${\rm 1s(^2S)2s2p^2(^2P)\ ^3P}$   & 66.4 & 55.5 & 58.5\tablenotemark{a}, 55.0\tablenotemark{b} \\
${\rm 1s2s^22p\ ^1P^o}$           & 55.2 & 48.3 & 53.8\tablenotemark{a}, 58\tablenotemark{c} \\
${\rm 1s2p^3\ ^1P^o}$             & 44.9 & 47.0 & 43.8\tablenotemark{a}, 43\tablenotemark{c} \\
${\rm 1s2p^3\ ^3D^o}$             & 80.7 & 80.3 & 75.0\tablenotemark{a}, 53.87\tablenotemark{d} \\
${\rm 1s2p^3\ ^3P^o}$             & 47.4 & 47.5 & 45.5\tablenotemark{a}, 34.16\tablenotemark{d} \\
\enddata
\tablenotetext{a}{MCDF method \citep{che87}}
\tablenotetext{b}{Breit--Pauli saddle-point complex-rotation method \citep{lin01}}
\tablenotetext{c}{Breit--Pauli saddle-point complex-rotation method \citep{lin02}}
\tablenotetext{d}{Breit--Pauli saddle-point complex-rotation method \citep{zha05}}

\end{deluxetable}


\begin{deluxetable}{rrrrrlrrr}
\tabletypesize{\scriptsize}
\tablecaption{Valence and Auger levels for nitrogen ions
\label{elec1}}
\tablewidth{0pt}
\tablehead{
\colhead{$N$} & \colhead{$i$} & \colhead{$2S+1$} & \colhead{$L$} & \colhead{$2J$} &
\colhead{Configuration} & \colhead{Energy} & \colhead{$A$r} & $A$a \\
\colhead{} & \colhead{} & \colhead{} & \colhead{} & \colhead{} &
\colhead{} & \colhead{eV} & \colhead{s$^{-1}$} & s$^{-1}$ \\
}
\startdata
 3 &  1 & 2 & 0 & 1 & 1s22s 2S1/2            &   0.0000 & $-$9.99E+02 & $-$9.99E+02 \\
 3 &  2 & 2 & 1 & 1 & 1s22p 2Po1/2           &   9.9459 &    3.32E+08 & $-$9.99E+02 \\
 3 &  3 & 2 & 1 & 3 & 1s22p 2Po3/2           &   9.9774 &    3.35E+08 & $-$9.99E+02 \\
 3 &  4 & 2 & 0 & 1 & 1s2s2 2S1/2            & 410.1562 &    1.22E+11 &    9.79E+13 \\
 3 &  5 & 4 & 1 & 1 & 1s(2S)2s2p(3Po) 4Po1/2 & 413.0288 &    5.05E+06 &    1.44E+07 \\
 3 &  6 & 4 & 1 & 3 & 1s(2S)2s2p(3Po) 4Po3/2 & 413.0471 &    1.27E+07 &    3.65E+07 \\
 3 &  7 & 4 & 1 & 5 & 1s(2S)2s2p(3Po) 4Po5/2 & 413.0778 &    0.00E+00 &    1.76E+07 \\
 3 &  8 & 2 & 1 & 1 & 1s(2S)2s2p(3Po) 2Po1/2 & 420.8755 &    1.65E+12 &    7.63E+12 \\
 3 &  9 & 2 & 1 & 3 & 1s(2S)2s2p(3Po) 2Po3/2 & 420.8968 &    1.65E+12 &    7.30E+12 \\
 3 & 10 & 4 & 1 & 1 & 1s(2S)2p2(3P) 4P1/2    & 424.3267 &    7.99E+08 &    5.54E+06 \\
 3 & 11 & 4 & 1 & 3 & 1s(2S)2p2(3P) 4P3/2    & 424.3446 &    8.06E+08 &    4.65E+08 \\
 3 & 12 & 4 & 1 & 5 & 1s(2S)2p2(3P) 4P5/2    & 424.3741 &    8.15E+08 &    2.85E+09 \\
 3 & 13 & 2 & 1 & 1 & 1s(2S)2s2p(1Po) 2Po1/2 & 425.6391 &    1.63E+11 &    6.06E+13 \\
 3 & 14 & 2 & 1 & 3 & 1s(2S)2s2p(1Po) 2Po3/2 & 425.6545 &    1.55E+11 &    6.10E+13 \\
 3 & 15 & 2 & 2 & 3 & 1s(2S)2p2(1D) 2D3/2    & 429.2436 &    8.33E+11 &    1.07E+14 \\
 3 & 16 & 2 & 2 & 5 & 1s(2S)2p2(1D) 2D5/2    & 429.2442 &    8.32E+11 &    1.07E+14 \\
 3 & 17 & 2 & 1 & 1 & 1s(2S)2p2(3P) 2P1/2    & 430.3236 &    2.68E+12 &    1.20E+08 \\
 3 & 18 & 2 & 1 & 3 & 1s(2S)2p2(3P) 2P3/2    & 430.3597 &    2.68E+12 &    4.50E+10 \\
 3 & 19 & 2 & 0 & 1 & 1s(2S)2p2(1S) 2S1/2    & 437.1874 &    7.86E+11 &    1.60E+13 \\
\enddata
\tablecomments{A complete machine-readable version of this table may be accessed on-line.}
\end{deluxetable}


\begin{deluxetable}{rrrrrr}
\tabletypesize{\scriptsize}
\tablecaption{Radiative K-transition data nitrogen ions
\label{elec2}}
\tablewidth{0pt}
\tablehead{
\colhead{$N$} & \colhead{$j$} & \colhead{$i$} & \colhead{Wavelength} & \colhead{$A$-coefficient} &
\colhead{$gf$-value} \\
\colhead{} & \colhead{} & \colhead{} & \colhead{0.1~nm} & \colhead{s$^{-1}$} &
\colhead{} \\
}
\startdata
 3 &   4 & 2 & 30.9798 & 4.08E$+$10 & 1.17E$-$02 \\
 3 &   4 & 3 & 30.9822 & 8.08E$+$10 & 2.33E$-$02 \\
 3 &   5 & 1 & 30.0183 & 5.05E$+$06 & 1.36E$-$06 \\
 3 &   6 & 1 & 30.0170 & 1.27E$+$07 & 6.87E$-$06 \\
 3 &   8 & 1 & 29.4586 & 1.65E$+$12 & 4.29E$-$01 \\
 3 &   9 & 1 & 29.4571 & 1.65E$+$12 & 8.61E$-$01 \\
 3 &  10 & 2 & 29.9203 & 9.70E$+$06 & 2.61E$-$06 \\
 3 &  10 & 3 & 29.9226 & 2.36E$+$05 & 6.34E$-$08 \\
 3 &  11 & 2 & 29.9191 & 7.01E$+$04 & 3.77E$-$08 \\
 3 &  11 & 3 & 29.9213 & 1.58E$+$07 & 8.49E$-$06 \\
 3 &  12 & 3 & 29.9192 & 2.14E$+$07 & 1.72E$-$05 \\
 3 &  13 & 1 & 29.1289 & 1.62E$+$11 & 4.12E$-$02 \\
 3 &  14 & 1 & 29.1279 & 1.54E$+$11 & 7.83E$-$02 \\
 3 &  15 & 2 & 29.5695 & 7.16E$+$11 & 3.76E$-$01 \\
 3 &  15 & 3 & 29.5717 & 1.17E$+$11 & 6.12E$-$02 \\
 3 &  16 & 3 & 29.5717 & 8.32E$+$11 & 6.55E$-$01 \\
 3 &  17 & 2 & 29.4935 & 1.79E$+$12 & 4.68E$-$01 \\
 3 &  17 & 3 & 29.4957 & 8.89E$+$11 & 2.32E$-$01 \\
 3 &  18 & 2 & 29.4910 & 4.24E$+$11 & 2.21E$-$01 \\
 3 &  18 & 3 & 29.4932 & 2.26E$+$12 & 1.18E$+$00 \\
 3 &  19 & 2 & 29.0197 & 2.56E$+$11 & 6.46E$-$02 \\
 3 &  19 & 3 & 29.0218 & 5.28E$+$11 & 1.33E$-$01 \\
\enddata
\tablecomments{A complete machine-readable version of this table may be accessed on-line.}
\end{deluxetable}


\end{document}